# Artificial Intelligence Models for Cell Type and Subtype Identification Based on Single-Cell RNA Sequencing Data in Vision Science

Yeganeh Madadi, Aboozar Monavarfeshani, Hao Chen, W. Daniel Stamer, Robert W. Williams, and Siamak Yousefi, Senior Member, IEEE

**Abstract**—Single-cell RNA sequencing (scRNA-seq) provides a high throughput, quantitative and unbiased framework for scientists in many research fields to identify and characterize cell types within heterogeneous cell populations from various tissues. However, scRNA-seq based identification of discrete cell-types is still labor intensive and depends on prior molecular knowledge. Artificial intelligence has provided faster, more accurate, and user-friendly approaches for cell-type identification. In this review, we discuss recent advances in cell-type identification methods using artificial intelligence techniques based on single-cell and single-nucleus RNA sequencing data in vision science.

—————————— ◆ ——————————

## 1 INTRODUCTION

SINGLE-cell RNA sequencing (scRNA-seq) technologies were established in 1990's [1]. These technologies provide a powerful way for researchers to efficiently distinguish phenotypic heterogeneity of complex cell populations types in biological samples [2].

Traditional scRNA-seq cell-type identification methods include two stages. First, the cells are clustered through an unsupervised approach then the clusters are annotated based on the canonical markers that are recognized in differentially expressed genes in each cluster [3-5]. In these methods, cluster annotation is very time-consuming because it requires broad literature knowledge about marker genes and forces calls whether to lump or split clusters based upon overlapping expression of marker genes.

In recent years, another class of models have been introduced to automatically identify cell types without the requirement of manual annotation. These models utilize annotated scRNA-seq atlases to train the models and identify the cell types of unlabeled data based on the labels from the cells in the atlases. Most of these models use feature learning based on annotated training data (e.g., related reference atlases). Feature learning includes a set of techniques that allows a model to automatically discover the representations appropriate for classification task from raw data. These approaches have a common purpose, which is accurate cell annotation; however, they differ in proposed algorithms and the usage of prior knowledge [6, 7]. These approaches are useful when the user has access to large and well annotated datasets (atlases) [7-9].

The scRNAseq technology has been used in numerous research domains including vision and ophthalmology. For example, scRNA-seq was applied to retinal tissues for cell identification as early as 2015 [10]. Shortly after, several scRNA-seq approaches were applied to various visual system including retina [11-28], outflow pathways [29-31], cornea [31-34], iris [31, 35], and ciliary body [31, 35, 36]. The eye can be used as a "window to the brain" as the structures of the eye are more accessible and can be more readily visualized using various imaging technologies compared to the brain proper. However, identifying eye's cell types is highly challenging due to the specialized/hybrid nature of many cells that facilitate vision. Additionally, the numerous cell types of the eye collectively make it a structure of interest for investigating different health conditions. The light enters the eye and reaches the retina by

————————————————

- *Yeganeh Madadi is with the Department of Ophthalmology, University of Tennessee Health Science Center, Memphis, TN, USA.*
  *E-mail: ymadadi@uthsc.edu.*
- *Aboozar Monavarfeshani is with the Center for Brain Science and Department of Molecular and Cellular Biology, Harvard University, Cambridge, MA. USA.*
  *F.M. Kirby Neurobiology Center, Boston Children's Hospital, Boston, MA, USA.*
  *E-mail: aboozar.monavarfeshani@childrens.harvard.edu.*
- *Hao Chen is with the Department of Pharmacology, Addiction Science and Toxicology, University of Tennessee Health Science Center, Memphis, TN, USA.*
  *E-mail: hchen@uthsc.edu.*
- *W. Daniel Stamer is with the Department of Ophthalmology, Duke Eye Center, Duke University, Durham, NC, USA.*
  *E-mail: william.stamer@duke.edu.*
- *Robert W. Williams is with Department of Genetics, Genomics, and Informatics, University of Tennessee Health Science Center, Memphis, TN, USA.*
  *E-mail: rwilliams@uthsc.edu.*
- *Siamak Yousefi is with the Department of Ophthalmology, University of Tennessee Health Science Center, Memphis, TN, USA.*
  *Department of Genetics, Genomics, and Informatics, University of Tennessee Health Science Center, Memphis, TN, USA.*
  *E-mail: siamak.yousefi@uthsc.edu.*

penetrating the cornea and lens, which in turn are responsible for focusing the light appropriately on retina. The muscles of the iris, located posterior to the lens and the anterior cornea, regulates the angle of light that can enter the eye. The ciliary body and trabecular meshwork tissues control intraocular pressure via the regulation of the production and drainage of the aqueous humor, respectively. The retinal pigment epithelial (RPE) cells are located adjacent to photoreceptor cells lining the back of the orbit and serves to recycle visual pigment and provide nutrients to retinal photoreceptors. All these ocular tissues have major clinical impacts: For instance, lens opacification leads to cataracts, elevated intraocular pressure (IOP) may lead to glaucoma, and the retinal pigment epithelium dysfunction may eventually lead to age-related macular degeneration (AMD). Based on the WHO report, approximately 33.6 million people worldwide are blind in the 50 years and older age group [37]. The major distinguished causes of blindness are cataracts (~15.2 million), glaucoma (~3.6 million), and age-related macular degeneration (AMD) (~1.8 million) [38]. Over the past few years, significant scRNA-seq data have been generated from different eye tissues using various single cell type and subtype identification approaches. While important and promising, these findings would be made even more useful and valuable by benchmarking different computational approaches to determine which method(s) is better suited to solve a particular problem or identify specific cell types. Additionally, a benchmarking review would aid investigators who are new to single cell and vision research choosing the most appropriate tools, thus avoiding potential confusion and the need to make selections based on trial and error.

In this paper, we first review the published scRNA-seq datasets (atlases) related to the visual system. We then review the computational approaches for broad cell type and subtype identification. Further, we provide several taxonomies based on datasets and well-established computational approaches. Finally, we benchmark different computational approaches based on three vision-related datasets, which are perfect for this type of analysis-not too big or small, and discuss advantages and disadvantages, as well as potential applications.

## 2 THE SCRNA-SEQ/SNRNA-SEQ ATLASES (DATASETS) OF THE VISUAL SYSTEM

### 2.1 Atlases of the mouse retina

Macosko et al. [10] were among the first groups who applied high-throughput scRNA-seq (i.e., Drop-seq technology) and generated scRNA-seq data from ocular tissues. They used unsupervised machine learning methods and identified 39 cell clusters from 44,808 single retinal cells. They annotated 33 cell types comprised of both rod and cone photoreceptors (PRs), horizontal cells (HCs), bipolar cells (BCs), amacrine cells (ACs), retinal ganglion cells (RGCs), Müller glia (MG), and non-neuronal cells. A follow-up study led by Shekhar [11] enriched mouse retinal bipolar cells (BCs) before subjecting them to the scRNA-seq. From 23,300 single-cell transcriptomes, they derived a molecular classification that identified 15 BC subtypes including all types identified previously by Macosko et al.

TABLE 1
OVERVIEW OF THE SCRNA-SEQ/SNRNA-SEQ ATLASES GENERATED BASED ON VISION SYSTEM

| Reference | Year | # of Cells | Species/Tissues | # of Clusters | Deposited Data |
|---|---|---|---|---|---|
| Macosko et al. [10] | 2015 | 44,808 | Mouse/Retina | 39 | GSE63473 |
| Shekhar et al. [11] | 2016 | 23,300 | Mouse/Retinal bipolar cells | 18 | GSE81905 |
| Rheaume et al. [12] | 2018 | 6225 | Mouse/Retinal ganglion cells | 41 | GSE115404 |
| Ariss et al. [13] | 2018 | 11,500 | Drosophila/Eye disc | 15 | GSE115476 |
| Tran et al. [14] | 2019 | 35,699 | Mouse/Retinal ganglion cells | 45 | GSE137400 |
| Peng et al. [15] | 2019 | 165,000 | Macaque/Fovea and peripheral retina | Fovea:64 Periphe:71 | GSE118480 |
| Heng et al. [16] | 2019 | 64,196 | Mouse/Retina | 12 | GSE132229 GSM3854512–3854519 |
| Menon et al. [17] | 2019 | 20,091 | Human/Retina | 9 | GSE137537 GSE137847 |
| Lukowski et al. [18] | 2019 | 20,009 | Human/Retina | 18 | E-MTAB-7316 |
| Voigt et al. [19] | 2019 | 8,217 | Human/Fovea and peripheral retina | 17 | GSE130636 |
| Liang et al. [20] | 2019 | 5873 | Human /Retina | 7 | GSE133707 |
| Yan et al. [21] | 2020 | 84,982 | Human/Fovea and peripheral retina | 58 | GSE148077 |
| Cowan et al. [22] | 2020 | 285,441 | Human/Fovea and periphery, pigment epithelium, and choroid | 65 | GSE104827 |
| Orozco et al. [23] | 2020 | 100,055 | Human/Retina and retinal pigment epithelium | 27 | GSE135092 GSE135133 |
| Yan et al. [24] | 2020 | 32,523 | Mouse/Retinal amacrine cells | 63 | GSE149715 |
| Patel et al. [29] | 2020 | 8,758 | Human/Outflow pathways | 12 | PRJNA616025 |
| Van Zyl et al. [30] | 2020 | 24,023 | Human, macaque, pig, and mouse/Outflow pathways | 19 | GSE146188 |
| Sun et al. [25] | 2021 | 14,424 | Mouse/Retina | 28 | GSE178121 |
| Yamagata et al. [26] | 2021 | 40,000 | Chick/Retina | 150 | GSE159107 |
| Yi et al. [27] | 2021 | 119,520 | Humans and macaques/Fovea and peripheral retina | 56 | GSA: CRA002680 GSA: HRA000182 |
| Kö̈lsch et al. [28] | 2021 | 32,679 | Zebrafish/Retinal ganglion cells | 32 | GSE152842 |
| Collin et al. [32] | 2021 | 21,343 | Human/Cornea | 21 | GSE155683 |
| Català et al. [33] | 2021 | 19,472 | Human/Cornea | 15 | GSE186433 |
| Wang et al. [35] | 2021 | 34,357 | Mouse/Iris and ciliary body | 10 | GSE183690 |
| Youkilis et al. [36] | 2021 | 10,024 | Mouse/Ciliary body and contiguous tissues | 22 | GSE178667 |
| Thomson et al. [39] | 2021 | 26,008 | Mouse/Outflow pathways | 25 | GSE168200 |
| Wang et al. [34] | 2022 | 16,924 | Human/Corneal endothelium | 4 | GSA: HRA000781 |
| Van Zyl et al. [31] | 2022 | 195,248 | Human/Anterior segment | 60 | GSE199013 |

[10] and two novel cell types. In total, they identified 18 BC clusters in this study. In another study, Rheaume et al. [11] generated 6225 RGCs from mouse eyes, developed a computational pipeline, and analyzed scRNA-seq data. They identified 40 clusters of cells corresponded to 40 different subtypes of RGCs in which 30 RGCs subtypes were recognized previously. They identified that most of the previously known RGC markers were expressed in multiple RGC clusters (subtypes) and only NPY, Jam2, Trhr, Pde1a, and Gna14 genes were expressed in a single cluster (RGC subtype). Therefore, they could determine the gene expression (variability) threshold required for RGC subtype segregation and diversification and presented a hierarchy from RGC cell types to subtypes. Further, they generated a portal for comparing gene expression in RGC subtypes. Tran et al. [14] in 2019 and Yan et al. [24] in 2020 adapted the pipeline introduced in Shekhar et al. [11] to classify a large number of RGCs and amacrine cells obtained from mouse retina. They identified 45 RGC subtypes based on an scRNA-seq dataset with 35,699 single cells and 63 subtypes of amacrine cells. Yan et al. [24] adopted a strategy of selective depletion from 55,287 single cells, of which 32,523 cells were identified as amacrine cells. They identified 63 subtypes of amacrine cells which led to identifying class-specific marker for amacrine cells. They identified molecular markers in each type and utilized them for the characterization of morphology in multiple types. They showed that these markers included all of the previously known AC types as well as numerous new markers. They found most of the AC types expressed markers in canonical inhibitory neurotransmitters glycine or GABA, but some of the markers expressed neither in glycine nor in GABA. They also found transcriptomic relationships among AC types and identified transcription factors that were expressed through individual or multiple closely related cell types. They observed that Tcf4 and Meis2 genes were expressed by most glycinergic and most GABAergic types, respectively.

Heng et al. [15] proposed an scRNA-seq study to identify spontaneous uveoretinitis in Aire−/− mice by utilizing 64,196 single retinal cells. They analyzed the patterns of gene expression in immune and retinal cells based on the abundance of different immune cell types and identified six different classes: Th1 cells, Cd8a+ T cells, T follicular Helper (Tfh) cells, regulatory T (Treg) cells, NK cells, and NK T cells. Sun et al. [24] constructed an atlas based on 14,424 single cells collected from healthy mice and mice with diabetes to identify pathological alterations of diabetic retinopathy (DR). They determined several pathological variations of DR and introduced potential guides for DR therapy.

However, while the mouse visual system provides an invaluable resource for studying the structure, function, and development of neural circuits for vision, it has significant shortcomings when compared to the human visual system. Essentially, retinas, vascular arrangement, and lamina of primates, such as humans, differ from those of rodents but some other structures such as conventional outflow system are very similar. Most notably, primates, unlike rodents, have a small, specialized central retinal region, the fovea, which is responsible for chromatic vision [40]. In order to address these differences shortcomings and gain a broad comprehensive understanding picture of specific mechanisms underlying cell heterogeneity in human vision retina, several groups have generated single cells from human and other primate visual systems retina.

## 2.2 Retina atlases in non-human and human primates

In 2019, Peng et al. [15] generated a retinal atlas from the macaque monkey, profiling the foveal and peripheral retina separately. In order to enhance the probability of being able to the identify rare RGCs types that are localized within the highly rod-dominated peripheral region, they enriched RGCs (utilizing anti-Thy1) for some peripheral samples and depleted rods (utilizing anti-CD73) from the others in order to certify the identification of rare types. From 165,000 single cells, they identified 64 foveal and 71 peripheral cell types. In several studies, macaque atlases have been used to annotate different clusters of retinal cells [17-23, 27]. Over the past three years, many research groups have extended related studies to foveal and peripheral retina of humans [19, 21, 22, 27].

Consistent with physiological and morphological studies, cell types in human and macaque retinas are highly similar. In 2019, Menon et al. [17] generated the first human retina single-cell transcriptomic atlas based on 20,091 single cells. Their analysis was based on a multi-resolution network to identify the retinal cell types and the corresponding gene expression signatures. They found that human retinal microglia are more heterogeneous than previously thought and recognized some cell types (glia, cone photoreceptors, and vascular cells) associated with the risk of age-related macular degeneration (AMD). Lukowski et al. [18] compiled a transcriptome atlas from human retina by profiling 20,009 single cells. They identified 18 transcriptionally distinct cell types representing all known neural retinal cells: retinal ganglion cells, Müller glia, bipolar cells, amacrine cells, rod photoreceptors, cone photoreceptors, horizontal cells, microglia, and astrocytes. Liang et al. [20] generated an atlas based on 5873 single nuclei collected from human retinal tissue. They identified the main retinal cell types and corresponding marker genes. They proposed the photoreceptor cell DEG list as a potential gene prioritization tool for new inherited retinal disease (IRD) gene discovery and showed that their atlas improves prioritization of genes associated of human retinal diseases in comparison to mouse snRNA-seq and human bulk RNA-seq data.

Several teams have generated atlases of single cells collected from human fovea and peripheral retina [19, 21-23, 27]. Voigt et al. [19] proposed an scRNA-seq for the human fovea and peripheral retina. They retrieved 8217 cells, including 3578 cells from fovea and 4639 cells from periphery. They identified two clusters of cone photoreceptors based on gene expression differences. One cone cluster included 96% of foveal cells, while the other cluster included exclusively peripherally- isolated cone cells. The major marker for peripheral cones was Beta-Carotene Oxygenase 2. They observed a relative defect caused by this enzyme that may be due to the carotenoids accumulation which is responsible for the yellow pigment in the macula.

The other study for understanding disease pathogenesis and identifying causal AMD genes was performed by Orozco et al. [23]. They presented an atlas of the human retina and retinal pigment epithelium (RPE) and comprised an expression quantitative trait loci (eQTL) atlas that included RPE/choroid and macula-specific retina, and single-nucleus RNA-seq from the human retina. They found enriched expression for AMD gene candidates in RPE cells. They identified 15 putative genes for AMD according to genetic association signals within AMD risk and eye eQTL, containing the genes TSPAN10 and TRPM1. Yan et al. [21] analyzed the transcriptomes of 84,982 cells of the fovea and peripheral retina and recognized 58 cell types including retinal ganglion, horizontal, bipolar, photoreceptor, amacrine, and non-neuronal cells. They found that all cell types are common between two retinal regions, however there are significant differences in gene expression and proportions among foveal and peripheral groups of shared types.

Cowan et al. [22] used human retinal tissues with multiple synaptic and nuclear layers for studying disease mechanisms and treatment in human retinas. They sequenced 285,441 single cells from the fovea and periphery, choroid, and pigment epithelium of human retinas. Comparing the periphery to the fovea, they recognized regional characteristics and features of cell types and, through comparing organoid to organ, they determined that organoid cell types transcriptomes converge to those of peripheral retinas in adult humans.

Yi et al. [27] in 2021 provided a transcriptomic atlas from 119,520 cells of the foveal and peripheral retina from humans and macaques of different ages. The molecular features in retinal cells differed between the two species, suggesting distinct regional and species specializations in human and macaque retinas. However, human retinal aging occurred in a cell-type-specific and region- manner. The aging of the human retina showed a foveal-to-peripheral gradient. Quantification analysis of scRNA-seq data showed that the population of MYO9A⁻ rods greatly decreased during aging of human retina.

### 2.3 Retina atlases in insects, avians, and fish

Ariss et al. [13] generated 11,500 single eye disc cells including all main known cell types and determined the impact of an Rbf mutation within Drosophila eye development. They identified a transcriptome occurring in differentiating photoreceptors during axonogenesis and identified a cell population that showed intracellular acidification due to growth of glycolytic activity. Yamagata et al. [26] utilized scRNA-seq for generating a chick retina cell atlas to study the avian visual system. They recognized 136 cell types and 14 developmental or positional intermediates distributed among six classes: retinal ganglion, amacrine, photoreceptor, horizontal, bipolar, and glial cells. They adopted an approach based on CRISPR technology, named eCHIKIN to annotate different cell types and to analyze selectively expressed genes to recognize the molecularly defined types of morphology. For Muller glia, they observed that transcriptionally different cells were regionally localized along the dorsal-ventral, anterior-posterior, and central-peripheral retinal axes. They also identified immature horizontal cells, photoreceptors, and oligodendrocyte types which persist into late embryonic stages. Kolsch et al. [28] systematically classified RGCs in adult and larval zebrafish, thereby identifying marker genes for more than 30 mature types and various developmental intermediates.

### 2.4 Atlases from other ocular tissues

Most scRNA-seq atlases for vision were generated based on retinal tissues. However, since 2020, several researchers have generated atlases based on other ocular tissues as well.

The outflow pathway for intraocular fluid (aqueous humor), comprises one of the complex tissues responsible for maintaining intraocular pressure (IOP) homeostasis. Dysfunction of these tissues' cell-types results in ocular hypertension and risk for glaucoma, which is a leading cause of blindness. Patel et al. [29] generated a scRNA-seq atlas from human outflow tissues. They obtained expression profiles from 8758 cells and identified 12 different cell types. A major utility of their atlas is to map glaucoma-relevant genes to the human outflow cell types. They also presented two different TM cell types, showed that SC is a hybrid blood lymphatic vessel, and highlighted the abundance of resident macrophages in the outflow tissues. Van Zyl et al. [30] performed similar studies of outflow tissues on humans as well as four other species including mouse (Mus musculus), cynomolgus macaque (Macaca fascicularis), rhesus macaque (Macaca mulatta), and pig (Sus scrofa). They analyzed scRNA-seq data from 24,023 single cells and identified 19 cell types. While they observed that many human cell types were similar to those from other species, they found that there were also some differences in both cell types and the marker genes. They observed that many human cell types had counterparts in other species, but differences in gene expression and cell types were evident between human and non-human tissues as well. They also identified cell types that expressed genes that were known to be associated with glaucoma. Thomson et al. [39] showed that tissue-specific elimination of Svep1 or Angpt1 from the trabecular meshwork caused primary congenital glaucoma in mouse with severe defects in the adjacent Schlemm's canal. By single-cell transcriptomic analysis on glaucomatous and normal Angpt1, they identified distinct trabecular meshwork and Schlemm's canal cell populations and discovered additional distinct trabecular meshwork-Schlemm's canal signaling pathways.

As light enters the eye, it passes through five layers of cornea including endothelium, epithelium, stroma, Bowman's layer, and Descemet's membrane. The cornea's structure and transparency formations are maintained by different cell-types populated in each layer. Attempts to understand disease conditions and to regenerate corneal tissue requires extensive knowledge for modifying cell profiles across this heterogeneous tissue. Collin et al. [32] analyzed 21,343 single cells collected from human corneas and adjacent conjunctivas and identified 21 clusters of cells. They extended their single cell study to keratoconus disease and showed that 1) activation of collagenase in corneal stroma and 2) reduced number of limbal suprabasal cells are key variations underlying the disease phenotype. Català et al. [33] generated a single cell transcriptomic atlas based on 19,472 single cells collected from corneal endothelium. They analyzed the corneal layers heterogeneity and identified HOMER3, CAV1, and CPVL expressions in the corneal epithelial limbal stem cell niche. They showed STMN1, CKS2, and UBE2C were individually expressed across highly proliferative transit-amplifying cells, NNMT was individually expressed through stromal keratocytes, and CXCL14 was expressed individually in suprabasal/superficial limbus. These results provide a basis for future amendments to current primary cell expansion protocols in order to enhance future profiling of corneal disease states. Wang et al. [34] introduced a transcriptomic atlas based on 16,924 single cells collected from human corneal endothelium. The corneal endothelium is a major tissue for maintaining corneal clarity through mediating hydration by pump and barrier functions. Their findings provide novel insights into the development of Fuchs endothelial corneal dystrophy (FECD) and suggest that NEAT1 may offer an attractive method for treating FECD.

The iris is responsible for controlling the level of retinal illumination via regulation of pupil diameter. Wang et al. [35] presented a study that provided snRNA-seq data generated from iris cells of mice. They identified major cell types for the mouse iris and the ciliary body, which led to the detection of two kinds of iris stromal cells and iris sphincter cells. They also characterized the diversities in cell-type transcriptomes within the dilated vs. resting states, then identified and validated antibody and in situ hybridization (ISH) probes to visualize the major iris cell types.

Youkilis et al. [36] introduced a transcriptomic atlas based on 10,024 cells collected from ciliary body of mice and contiguous tissues, which play. a major role in ocular homeostasis. They utilized scRNA-seq to assess the transcriptional signatures from the ciliary body and adjacent tissues. They identified two fibroblast signatures in the ciliary body cells (sclera and uvea), which were subsequently confirmed via in situ hybridization (ISH).

Van Zyl et al. [31] generated a single nucleus atlas of ocular anterior segment of the human eye. The anterior segment of the human eye includes cornea, iris, crystalline lens, ciliary body, and aqueous humor outflow pathways. They profiled 195,248 nuclei from anterior segment tissues and identified more than 60 cell types.

Table 1 shows the overview of the scRNA-seq and snRNA-seq atlases generated based on vision system. It is worth mentioning that the Spectacle web tool houses all single cell atlases for vision [41].

## 3 COMPUTATIONAL CELL-TYPES IDENTIFICATION METHODS FOR SCRNA-SEQ/SNRNA-SEQ

Table 2 presents a taxonomy for computational methods applied on scRNA-seq and snRNA-seq data to identify cell-types and subtypes based machine learning approaches. In this taxonomy, we have classified models based on how they learn to find different clusters or classes. According to this issue, we classified them into three main categories: unsupervised, semi-supervised, and supervised approaches. We have then provided sub-categories at each level and provided the references to those approaches.

### 3.1 Unsupervised methods for cell type and subtype identification

Unsupervised cell-type identification methods recognize cell types and subtypes based on different clustering techniques. In unsupervised learning models, the data has no labels (cells have no type or subtype labels). Labeling the data means annotating the samples into different categories. Unsupervised learning models identify different clusters (groups) of samples based on (interesting) structures or patterns in the data. The labels will be propagated to the samples once the clusters were identified. Clusters (cell types or subtypes here) must be validated as there is no prior knowledge regarding the true labels.

Unsupervised models can be further categorized to four subgroups including graph-based, partition-based, density-based, and deep neural network-based clustering.

TABLE 2
TAXONOMY OF CELL-TYPE IDENTIFICATION METHODS THAT WERE INCLUDED IN THIS STUDY

| | | | |
|---|---|---|---|
| Unsupervised | Graph-based | Nearest neighbor | Scmap [42], Seurat [4], [10], [11], [13], [14], [15], [17], [18], [19], [20], [22], [29], [25], [26], [27], [28], [32], [33], [21], [23], [24], [35], [36], [34], [30], scType [43] |
| | Partition-based | K-means clustering | [16] |
| | Density-based | DBSCAN clustering | [12] |
| | Deep NN-based | Recurrent network | ScScope [44] |
| | | Autoencoder | DESC [45], ScAIDE [46], scETM [47] |
| | | Hierarchical Bayesian | scVI [48] |
| Semi-supervised | Deep NN-based | Autoencoder | DISC [49], ScDCC [50] |
| Supervised | Similarity-based | | ScLearn [51] |
| | General classifier-based | XGBoost | CaSTLe [52] |
| | | Logistic regression | SCCAF [53] |
| | | Fisher's linear discriminant analysis | ScID [54] |
| | Deep NN-based | Capsule networks | ScCapsNet [55] |
| | | Weighted GNN | ScDeepSort [56] |
| | | Hierarchical FFNN | NeuCA [57] |
| | Transfer learning-based | | ItClust [58], SCTL [59] |

We discuss some of these approaches below.

Scmap [42] is a graph-based clustering method that measures the maximum similarity between reference data and query data to identify cell types. The optimum projection of a new query cell onto a reference data set is identified by the nearest neighbor approach. Seurat 3.0 [4] is a widely used toolkit in single-cell studies that uses this approach. Seurat essentially utilizes canonical correlation analysis (CCA) [3] and mutual nearest neighbors (MNNs) [60] to identify shared subpopulations across datasets. Seurat first jointly reduces the dimensionality of the reference and query datasets utilizing diagonalized CCA, then applies L2-normalization for the canonical correlation vectors. The model then searches MNNs in the common low-dimensional representation. Seurat encodes cellular relationships among datasets, which serve as the basis in all subsequent integration analyses. In other words, Scmap learns cell-type-specific gene expression information in the reference dataset only and ignores useful information in the query dataset. As such, the model could be vulnerable to the batch effect if reference and query datasets are generated in different batches. Although Seurat 3.0 uses information through the identification of anchor pairs in the reference and query datasets, it does not specifically use cell type information in the reference data. Many studies [10, 13, 15, 18-20, 25-29, 32-36] have used Seurat to perform clustering as part of their computational pipeline to identify cell types and subtypes. In the scType [43] approach, the authors have introduced a marker database as well as a cell-type identification approach based on the Louvain algorithm for unsupervised cell-type annotation. This approach provides an end-to-end pipeline for identifying single cell types and subtypes based on a built-in marker database. It is worth mentioning that Yan et al. [21, 24] and Van Zyl et al. [30] performed unsupervised clustering based on Louvain algorithm with Jaccard correction while Shekhar et al. [11], Tran et al. [14], and Cowan et al. [22] employed Infomap algorithm [61] in their graph-based unsupervised clustering approach.

Some of the approaches including Heng and colleagues [16] have used conventional unsupervised clustering models such as k-means. Based on this model, cell types and subtypes are identified by cross-referencing the clusters for expression of multiple known cell type-specific markers.

Rheaume et al. [12] used a density-based clustering approach, namely, DBSCAN [62] to identify cell types and subtypes. DBSCAN is an unsupervised clustering method that identifies distinctive clusters in data based on groups with high adjacent density of samples, while separated from other clusters with low adjacent density of samples.

scScope [44] is a deep learning-based cell-types identification method, which utilizes a recurrent network layer for performing imputations iteratively on entries with zero-valued of input scRNA-seq data. scScope model lets imputed output to be iteratively amended through a selected number of recurrent steps. If number of recurrent steps is equal to 1, the model reduces to a standard autoencoder [63].

DESC [45] is another unsupervised deep learning algorithm that iteratively learns gene expression representations that are specific for each cluster, and also learns cluster assignments in scRNA-seq data analysis. In other words, DESC initializes different parameters obtained via an autoencoder and then learns a nonlinear function for mapping the original scRNA-seq data space into the low dimensional feature space through optimizing the clustering objective function, iteratively. By performing this iterative procedure, each cell moves to its nearest cluster centroid and balances technical and biological differences among clusters, and gradually decreases the influence of batch effect. ScAIDE [46] is another unsupervised deep learning clustering method that utilizes the autoencoder imputation network along with the distance-preserved embedding network (AIDE) for learning the data representations and then uses random projection hashing based on the k-means algorithm (RPH-kmeans) for accommodating the identification of rare cell types. There are several mod-els based on neural networks and deep learning approaches. Single-cell Embedded Topic Model (scETM) [47] is a neural network-based autoencoder that has a linear decoder along with matrix tri-factorization. This method simultaneously learns topic embeddings, highly expressed gene embeddings, batch-effect linear intercepts, and encoder network parameters from scRNA-seq data. This method synthesizes existing pathway information with gene embeddings at training the model to further enhance interoperability by tri-factorizing the cells-genes matrix into cells-by-topics, topics-by-embeddings, and embeddings-by-genes.

Single-cell variational inference (scVI) [48] is a deep neural network along with a stochastic optimization function for obtaining the probabilistic representation as well as analyzing the gene expression in single cells data.

scVI analysis is performed based on the hierarchical Bayesian [64] along with conditional distributions specified through deep neural networks. scRNA-seq data is encoded by a nonlinear mapping function and is projected on a low-dimensional feature space with the latent vector of normal random variables. This latent representation is then decoded via another nonlinear mapping function to produce a posterior estimate for the distributional parameters of genes from each cell.

### 3.2 Semi-supervised methods for cell type and subtype identification

Approaches that use a combination of supervised and unsupervised learning for cell type and subtype identification are categorized in semi-supervised learning (SSL) group.

Semi-supervised learning methods are promising when a few labeled data are available thus allowing unsupervised components to complement their training with unlabeled data [65].

SSL approaches may address dropout problems in single cell data. Dropout is essentially one of the main challenges in scRNA-seq data analysis. Dropouts, the extra false zero expressions, causes a skewed distribution of gene expressions and result in the misclassification of cell types [66]. Recent advances in combinatorial indexing-based or droplet-based sequencing technologies have significantly grown the throughput from thousands to more than a million cells for a single experiment and caused more intense dropout problems because of shallow sequencing depth in per cell [67]. In the following, we will discuss a few SSL-based cell identification methods.

DISC [49] is an autoencoder-based semi-supervised deep learning network method to infer gene structure and expression of single cell data generated based on the dropout technology. Semi-supervised learning here generates a reliable imputation approach via learning information from zero-count genes and positive genes, that can be treated as unlabeled and labeled data, respectively. DISC works well when the information provided by labeled data is limited. Single Cell Deep Constrained Clustering (scDCC) [50] is another semi-supervised clustering method based on autoencoder that encodes prior knowledge in constraint information, which is integrated into the clustering algorithm via a loss function. Also, scDCC integrates domain knowledge for the clustering step.

### 3.3 Supervised methods for cell -type and subtype identification

Different approaches in this category typically utilize a previously annotated dataset as a reference to train supervised machine learning classifiers and use the trained model to identify cell types and subtypes from unlabeled data. However, in such supervised methods, it is expected that the reference and query datasets resemble each other, which is not often the case. This poses challenges in successful label propagation [60].

scLearn [51] is a supervised cell type identification method designed via measuring the similarity among query cells and each reference cluster centroid through thresholds and measurements learned from reference datasets. scLearn has the ability of identifying novel cell types which are absent in the reference datasets. It uses a multilabel single-cell assignment strategy to assign a single cell to proper time status and cell type simultaneously.

General classifier-based methods are another subgroup of supervised cell type identification methods. CaSTLe [52], classification of single cells by transfer learning, is a supervised method based on the XGBoost classifier [68]. This method selects genes with mutual information gain and genes with top mean expression, then removes correlated genes before classification. All these steps are performed to ensure the reference and query datasets are brought into a common denominator to perform an accurate transfer of the classification model. SCCAF [53] presents a self-projection method based on linear regression [69], and ScID (Single Cell IDentification) [54] utilizes the Fisher's Linear Discriminant Analysis [70] to recognize transcriptionally related cell types among scRNA-seq datasets. This method extracts markers through the reference dataset and then weighs their relevance on the target dataset via learning a classifier based on the putative population of different cells either expressing or not expressing these genes.

scCapsNet [55] and scDeepSort [56] are a subgroup of supervised deep neural network-based methods in proposed taxonomy. scCapsNet [55] has an interpretable deep-learning architecture utilizing capsule networks (scCapsNet). Capsule structure is a neuron vector that represents the properties of a specific object and captures hi-erarchical relations. scCapsNet creates the decision-making black box transparent via analyzing internal weight parameters between capsule structures. scDeepSort [56] is a pre-trained cell-type identification approach based on weighted graph neural networks (GNN) model. GNN is one of the popularly-used deep learning methods [71] that captures graph dependency via message passing among the graph nodes and keeps a state which represents required information from its neighborhood with ideal depth [72]. NeuCA [57] obtains the mean gene expression profile for each cell type and calculates the correlation matrix among cell types. If the correlation matrix includes highly correlated cell types, then the model constructs a tree structure via hierarchical feed-forward clustering and trains a series of neural networks based on this tree structure.

ItClust [58] is a supervised cell type classification method that utilizes external well-annotated reference data to gain popularity over unsupervised clustering algorithms. As expected, the performance of supervised methods is highly related to the reference data quality. To lower the dependency of the model on the reference dataset quality, ItClust employs a transfer learning approach [73] that takes advantage of cell type-specific gene expression information learned from a reference dataset, to help classify cell types on a newly generated query dataset. In other words, this method extracts related information from the reference dataset by considering genes which are highly variable, based on the reference dataset. This ensures that transferred expression patterns will be useful for dividing cell types based on the new datasets. SCTL [59] is another deep transfer learning method to detect cell types and subtypes in the scRNA-seq data. SCTL includes four loss functions where one loss function corresponds to domain adaptation and three loss functions correspond to the adversarial network. These loss functions eventually minimize the classification error.

## 4 ANALYSIS PIPELINE AND VISUALIZING SCRNA-SEQ/SNRNA-SEQ RESULTS

In this section, we discuss key steps in scRNA-seq data visualization. The gene expression data are analyzed using statistical and machine learning techniques to obtain intuition for cell-type heterogeneity in the transcriptomic level. These techniques include normalization, scaling, dimension reduction, clustering, and classification. However, visualization plays an important role in both qualitative and quantitative scRNA-seq data analysis (see Figure 1).

There are several different techniques for visualizing scRNA-seq and snRNA-seq results (Figure 1 A-F). The t-distributed stochastic neighbor embedding (t-SNE) is a widely used statistical approach to visualize high-dimensional data by assigning each data point to a location on a low-dimensional (i.e., two-dimensional) map. Figure 1A shows clusters of the molecular diversity of cells using tSNE on a two-dimensional map [74]. Each point corresponds to a cell, with transcriptionally similar cells being mapped to different groups. Cells are typically colored differently based on their defined cluster identity. Figure 1B shows the feature plot of cells in panel (A) at which cells are colored based on their level of expression of genes of interest (i.e., genes that were differentially expressed among clusters). Uniform manifold approximation and projection (UMAP) is similar to t-SNE. Both t-SNE and UMAP build a graph which represents data into high dimensional space and reconstruct the graph into a lower dimensional space. While t-SNE utilizes a Gaussian probability function to identify cell's neighbor, UMAP uses a fuzzy function to build neighborhood. Figure 1C shows violin plot of the expression levels (y axis) of single cells in Panel A based on a specific gene (i.e., a gene that was differentially expressed in cluster number 4 compared to other clusters). Each violin represents the distribution of the expression levels of cells within that particular cluster. Box and whisker plots within each violin correspond to the median value (black horizontal line), interquartile range (bars), range (vertical lines), and outliers (dots). Figure 1D shows a sample dot plot visualizing the expression pattern (both number of cells expressing a gene and the level of expression) of a set of genes across different clusters (x-axis). The size of each circle indicates the percentage of cells in the cluster expressing the gene, and the color saturation represents the normalized expression level. Along with the dot-plot, a dendrogram calculated using hierarchical clustering on top typically represents transcriptional interrelationships among clusters. Figure 1E shows a sample heat map plot. This ordered heatmap plot represents the expression patterns of specific marker genes in single cells grouped by their cells in different clusters. Figure 1F shows a sample confusion matrix plot representing transcriptional correspondence between clusters in two different datasets. Colors indicate the proportion of cells of a given cluster in one dataset assigned to a corresponding cluster in the other dataset in which the classification algorithm trained on cells in one of the datasets.

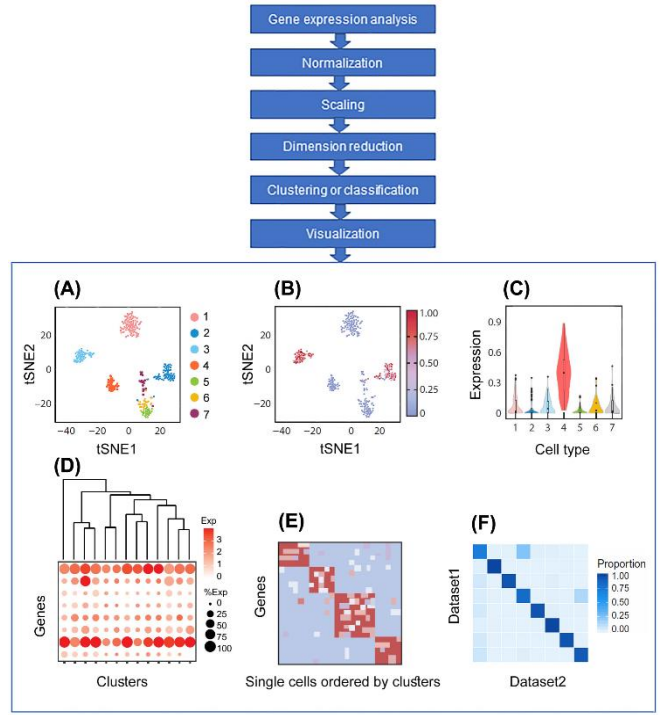

Fig. 1. Analysis pipeline and visualizing scRNA-seq and snRNA-seq results. (A) Cluster plot of two-dimensional visualization of the molecular diversity of cells using tSNE. (B) Feature plot with cells colored by their expression level of a single differential expression gene. (C) Violin plot where each violin is a representation of the probability distribution of the expression levels within each cluster. (D) Dot plot shows expression patterns of genes across a set of transcriptomically defined cell clusters. (E) Heat map plot shows expression patterns of specific marker genes in single cells grouped by their cell clusters. (F) Confusion matrix plot displays transcriptional correspondence between groups of cell types.

## 5 IDENTIFICATIONS OF OCULAR DISEASES BASED ON SCRNA-SEQ/SNRNA-SEQ DATA

In this section, we discuss ocular diseases and related tissues which have been studied on single-cell and single-nucleus RNA sequencing data. Most irreversible vision losses are caused by retinal diseases that explain the focus of most research on retina. The three prevalent and irreversible leading causes of vision loss are diabetic retinopathy, age-related macular degeneration (AMD), and glaucoma. Based on a single cell study, Lukowski et al. [18] observed that MALAT1, a long non-coding RNA, plays a major role in retinal homeostasis and disease [75] and they recognized that MIO-M1 cells have high levels of the thymosin beta 4 gene (TMSB4X) that has been related to glioma malignancy [76], and the calcyclin gene (S100A6), that is implicated in cone-associated or macular diseases [77]. These results show the differences and similarities in MIO-M1 for human retinal glial cells. Many groups have now utilized scRNA-seq and snRNA-seq data to assess retinal cell types which express genes implicated in different retinal diseases [15, 21-23, 27]. For example, most genes that are mutated in retinitis pigmentosa and initially affect rods, are selectively expressed via rods or via retinal pigment epithelial cells and required for rod viability. Also, most genes mutated in autosomal dominant optic atrophies and Leber hereditary optic neuropathies, which result in RGC death, are selectively expressed in RGCs; and many susceptibility genes related to diabetic retinopathy are expressed in vascular endothelial cells. There are various studies on AMD pathogenesis. CFH and ARMS2/HTRA1 genes impart biggest risk in AMD disease [78], Voigt et al. [19] evaluated the differentially expressed genes among cells of foveal versus peripheral origin.

Menon et al. [17] identified various cell types associated with AMD. Their results suggest that the genetic risk variants related to AMD affect cone photoreceptors, and they emphasized the importance of vascular and glial cells for disease pathogenesis. They found that expression of COL4A3, HTRA1, and vascular endothelial growth factor (VEGFA) had high scores for leading to AMD. Orozco et al. [22] identified enriched expression of AMD candidate genes for RPE cells. They identified TSPAN10 and TRPM1 which were enriched in retinal pigment epithelium, as causal genes that have high impact in early AMD disease. Yi et al. [27] reported that genes related to AMD are highly enriched in cones and foveal MG, suggesting a relation of regional cell subtype with this disease. Lyu et al. [79] showed that compositional changes are more pronounced in the macula in rods, endothelium, microglia, astrocytes, and Müller glia in the transition from normal to advanced AMD. They also identified enrichment in coagulation and complement pathways, signaling pathways, tissue remodeling, and antigen presentation, including PI3K-Akt, Rap1, Toll-like, and NOD-like. Sun et al. [25] studied diabetic retinopathy and identified four stress-inducible genes Rmb3, Cirbp, Mt1, and Mt2 which commonly exist in most retinal cell types. Diabetes increases the inflammatory factor gene expressions in retinal microglia and stimulates the immediate early gene expressions (IEGs) in retinal astrocytes. Van Zyl et al. [30] studied glaucoma cases and identified the cell types that represent gene expressions implicated in glaucoma. They found that several genes, such as MYOC, PITX2, CYP1B1, Cav1, and Cav2 are implicated in glaucoma. In their other study [31], they showed MYOC, ANGPT1, LMX1B, ANGPT2, PITX2, LTBP2, FOXC1, and CPAMD8 were associated with glaucoma. Patel et al. [29] studied assigning glaucoma-relevant genes to outflow cell clusters. They found that MYOC, PDPN, ANGPTL7, CHI3L1, and ANGPT1 were highly expressed in TM1 and TM2 cell types, while CAV1, CAV2, Tie2 (TEK), NOS3, ANGPT2, and PLAT, were highly expressed in vascular endothelial and lymphatic-like cell types.

In Figure 2, we show the percentage of scRNA-seq/snRNA-seq studies based on different ocular diseases and ocular tissues from different species We found that in vision and ophthalmology, mouse, human, macaque, and chick were studied the most (55%, 19%, 11%, 9%, respectively), followed by drosophila, pig, and zebrafish that were studied similarly (about 2% each).
From mouse studies, about 89%, 7%, 4% are focused on retina, iris, and outflow pathways tissues, respectively. It is worth mentioning that almost all of the studies on pig have only been applied to the outflow pathways tissues. From all the studies performed on macaque, about 82% are on retina and 18% on outflow pathways tissues. Almost all zebrafish studies have only been conducted on retina.

From studies done on human tissues, about 90% have analyzed retina and the remaining have focused on the outflow pathways tissues, respectively. From the studies performed on chick, about 22% were performed on retina and 78%, were applied on cornea (see Figure 2A). Also, we observed that most single cell studies have focused on AMD, diabetic retinopathy, glaucoma, and uveoretinitis. About 45% of the single cell studies were focused on AMD, whereas diabetic retinopathy, glaucoma, and uveoretinitis constituted 22%, 22%, and 11% of studies, respectively (see Figure 2B).

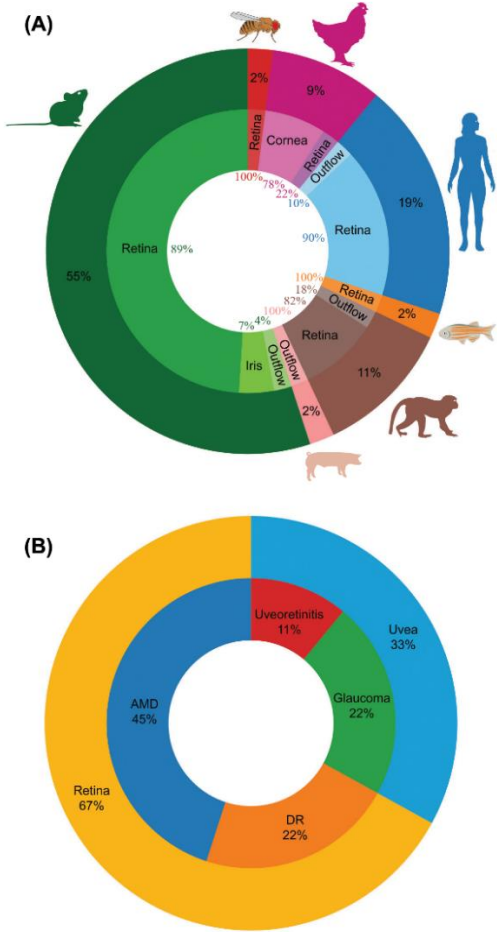

Fig. 2. Percentage of single cell studies in vision and ophthalmology based on different species, tissues, and diseases. (A) Percentage of studies conducted on mouse, pig, macaque, zebrafish, human, chick, drosophila were 55%, 2%, 11%, 2%, 19%, 9%, and 2%, respectively. From studies done on mouse species, about 89%, 7%, 4% are on retina, iris, and outflow pathways tissues, respectively. Almost all pig studies have only been done on outflow pathways tissues. From studies performed on macaque, 82% were on retina and 18% on outflow pathways tissues. Almost all zebrafish and drosophila studies have only been done on retina. From studies conducted on human, about 90% are on retina while about 10% are focused on outflow pathways tissues. From studies on chick, about 22% are focused on retina while nearly 78% are focused on cornea. (B) The percentage of studies performed on different diseases include about 45% on age related macular degeneration (AMD), 22% on diabetic retinopathy (DR), 22% on glaucoma, and 11% on uveoretinitis.

## 6 BENCHMARKING CELL-TYPE IDENTIFICATION METHODS FOR SCRNA-SEQ/SNRNA-SEQ DATA

### 6.1 Benchmark Datasets

Benchmarking could aid investigators who are new to single cell and vision research to select the most appropriate tools and validation datasets thus avoiding potential confusion and the need to perform trial and error. We introduced numerous scRNA-seq atlases in vision science and now benchmark numerous methods based on several of those atlases to highlight the advantage and limitations of those methods.

The datasets used for benchmarking vary in the number of cells, genes, and cell populations, thus resulting in different levels of challenges in identification of each cell types (see Table 3).

The first dataset, selected from the Macosko et al. [10] study, included 44,808 single cells from 39 different retinal cell types (and subtypes) with 24,658 common genes based on Droplet-based RNA-seq technology. Single cells in this dataset were isolated from retinas of 14-day-old wild-type C57BL/6 mice, which are genetically identical (within each strain) and thus exhibit a high degree of uniformity in their inherited characteristics and response to experimental treatments. Experimental workflow of this study was performed using seven different batches generating 3226, 6020, 8336, 5683, 6991, 6971, and 7581 cells (total 44,808 cells).

The second dataset, selected from the Rheaume et al. [12] study, included 6225 single RGCs from 41 different subtypes with 13,616 common genes. Cells were isolated from retinas of eight postnatal (day five; both sexes) C57Bl/6 mice in a single batch based on Droplet-based RNA-seq technology.

The third dataset, acquired from the Tran et al. [14] study comprised of 35,699 single RGCs in 45 different subtypes and 18,222 common genes which were collected from C57BL/6 adult mice retina from 6 to 20 weeks ages. Three batches were used to generate single RGCs based on droplet-based RNA-seq technology. Characteristics of the cells and cell types are shown in Table 3.

We randomly split the cells in these datasets into two subsets, where 80% of cells were allocated to the training data and the remaining 20% of cells were allocated to the testing data. After five rounds of cross-validation, the mean values of accuracy were used to evaluate the performance of the cell-types identification methods.

### 6.2 Data preprocessing

The count matrix and manually annotated labels were downloaded from public resources (GSE63473, GSE115404, GSE137400). We used the Seurat [4] R package (version 4.0.6) to normalize and scale the count transcriptome data. We excluded cells with fewer than 200 expressed genes and removed genes that were expressed in

TABLE 3
THE DATASETS USED FOR BENCHMARKING IN THIS STUDY

| Macosko et al. [10] | | Rheaume et al. [12] | | Tran et al. [14] | |
|---|---|---|---|---|---|
| Cluster name | #of Cells | Cluster name | #of Cells | Cluster name | #of Cells |
| HC | 252 | P5 RGC0 | 161 | W3-like1: RGC1 | 3,000 |
| RGC | 432 | W3D1: P5 RGC1 | 426 | W3D1: RGC2 | 2,859 |
| AC | 289 | P5 RGC2 | 188 | F-mini-ON: RGC3 | 1,990 |
| AC | 73 | P5 RGC3 | 196 | F-mini-OFF: RGC4 | 1,868 |
| AC | 77 | F-mini-ON: P5 RGC4 | 329 | J-RGC: RGC5 | 1,715 |
| AC | 211 | P5 RGC5 | 66 | W3B: RGC6 | 1,590 |
| AC | 326 | P5 RGC6 | 52 | RGC7 | 1,579 |
| AC | 159 | F-mini-OFF: P5 RGC7 | 268 | RGC8 | 1,258 |
| AC | 350 | F-RGC: P5 RGC8 | 95 | T-RGC: RGC9 | 1,223 |
| AC | 191 | P5 RGC9 | 185 | RGC10 | 1,170 |
| AC | 214 | P5 RGC10 | 143 | RGC11 | 990 |
| AC | 274 | P5 RGC11 | 88 | ooDSGC-N: RGC12 | 953 |
| AC | 50 | P5 RGC12 | 135 | W3-like2: RGC13 | 943 |
| AC | 111 | W3-like1: P5 RGC13 | 429 | RGC14 | 875 |
| AC | 73 | J-RGC: P5 RGC14 | 235 | RGC15 | 865 |
| AC | 262 | T-RGC-S2: P5 RGC15 | 91 | ooDSGC-D/V: RGC16 | 829 |
| AC | 375 | P5 RGC16 | 135 | T-RC-S1: RGC17 | 828 |
| AC | 83 | P5 RGC17 | 121 | RGC18 | 826 |
| AC | 127 | W3D3: P5 RGC18 | 80 | RGC19 | 775 |
| AC | 389 | P5 RGC19 | 115 | RGC20 | 711 |
| AC | 254 | P5 RGC20 | 224 | T-RGC-S2: RGC21 | 687 |
| AC | 274 | P5 RGC21 | 102 | MX: RGC22 | 610 |
| AC | 264 | P5 RGC22 | 48 | W3D2: RGC23 | 601 |
| Rods | 29,400 | P5 RGC23 | 93 | RGC24 | 553 |
| Cones | 1,868 | P5 RGC24 | 89 | RGC25 | 542 |
| BC | 2,217 | P5 RGC25 | 175 | RGC26 | 534 |
| BC | 664 | P5 RGC26 | 233 | RGC27 | 529 |
| BC | 496 | W3-like2: P5 RGC27 | 147 | F-midi-OFF: RGC28 | 517 |
| BC | 591 | P5 RGC28 | 100 | RGC29 | 499 |
| BC | 636 | P5 RGC29 | 124 | W3D3: RGC30 | 491 |
| BC | 512 | P5 RGC30 | 186 | M2: RGC31 | 444 |
| BC | 320 | ooDSGC-N: P5 RGC31 | 168 | F-RGC: RGC32 | 407 |
| BC | 849 | P5 RGC32 | 133 | M1a: RGC33 | 323 |
| M | 1,624 | P5 RGC33 | 108 | RGC34 | 312 |
| A | 54 | W3B: P5 RGC34 | 155 | RGC35 | 310 |
| F | 85 | P5 RGC35 | 70 | RGC36 | 236 |
| V | 252 | P5 RGC36 | 183 | RGC37 | 213 |
| P | 63 | P5 RGC37 | 135 | F-midi-ON: RGC38 | 207 |
| M | 67 | P5 RGC38 | 150 | RGC39 | 202 |
| | | P5 RGC39 | 44 | M1b: RGC40 | 174 |
| | | P5 RGC40 | 20 | aON-T: RGC41 | 126 |
| | | | | aOFF-S: RGC42 | 113 |
| | | | | aON-S/M4: RGC43 | 106 |
| | | | | RGC44 | 62 |
| | | | | aOFF-T: RGC45 | 54 |

HC = Horrizontal Cells, RGC = Retinal Ganglion Cells, AC = Amacrine Cells; BC = Bipolar Cells, M = Muller Glia, A = Astrocytes, F = Fibroblasts, V = Vascular Endothelium, P = Pericytes, M = Microglia.

fewer than 3 cells (based on Seurat). Then we selected highly variable genes. We computed the dispersion (variance/mean) and also the mean expression of each gene [10]. We then put the dispersion measure for each gene between all cells into several bins based on their average expression. Within each bin, the z-normalized of dispersion measure of all genes was calculated to identify highly variable genes. Based on Macosko et al. [10], the z-score cutoff was 1.7, which is the default parameter in Seurat. We selected a z-score of 1.7 which includes about 2000 highly variable genes.

### 6.3 Benchmark methods

Among the forty-five methods discussed in this review, we included seventeen published cell identification methods that their source codes or program was publicly accessible.

These methods are divided into three main categories: (1) Unsupervised cell-types identification methods, which are based on unsupervised clustering of the transcriptomes [4, 42, 45-48], (2) Semi-supervised cell-types identification methods, which require some limited labeled data in order to allow models to complement their training with unlabeled data [49, 50], (3) Supervised cell-types identification methods, which require a training dataset labeled with the corresponding cell populations to train the classifier [51-59].

### 6.4 Evaluation indicators

We evaluated the performance based on accuracy, speed, and memory usage metrics. Accuracy is defined as the ratio of cells that are correctly predicted cells, divided by the total number of annotated cells [9]. The memory requirements were obtained via reading rss (resident set size) attribute (which is returned through calling Process().memory_info() of psutil in the Python package) [80]. For the other approaches based on R language, we used the reticulate package [81] for calling the mentioned Python function for having consistency.

### 6.5 Results

#### 6.6.1 Accuracy of cell type identification

We compared seventeen cell-type identification methods for correctly identifying the number of cell types through applying each method on three ocular datasets [10, 12, 14] that contain 39, 41, and 45 cell types. Figure 3 illustrates the accuracy of these models applied on three different datasets.

Figure 3 presents the performance of different cell-types identification methods based on three ocular datasets. Macosko et al. [10] dataset contains heterogenous numbers of cells in different cell types and also includes single cells from seven different batches compared to the Tran et al. [14] and Rheaume et al. [12] datasets. The benchmarking results on Macosko et al. dataset showed that ItClust and SCTL generate the highest accuracy in comparison to the other methods. Both methods use a transfer learning-based algorithm, which improve cell-type identification particularly on datasets with varying number of cells from different cell types and removes the batch effect better than the other methods. In benchmarking the methods based on the Rheaume et al. and Tran et al. datasets, with over 40 cell types, NeuCA, ItClust, and SCTL outperformed the other methods. Generally, the results on all three datasets show that supervised cell identification methods obtained higher accuracy. In average, these three methods, ItClust, NeuCA, and SCTL, obtained the highest accuracy based on all three datasets.

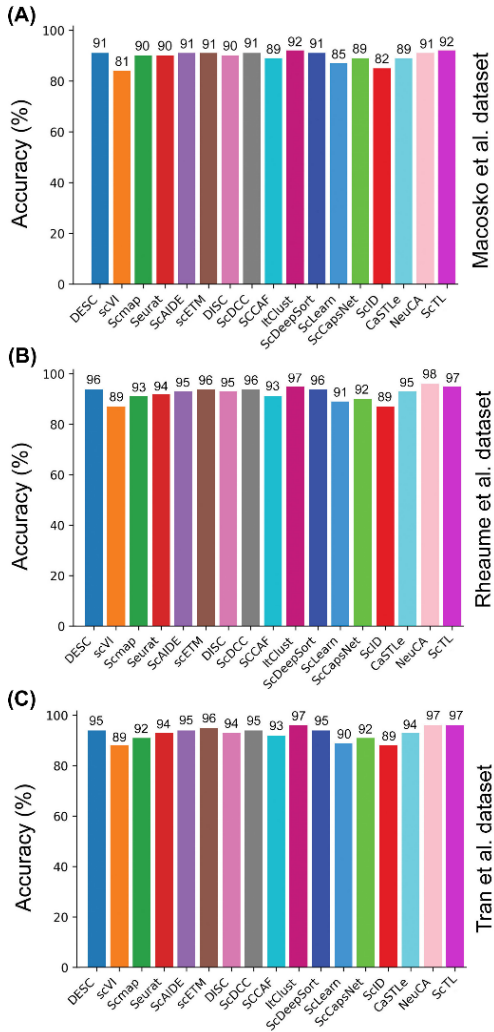
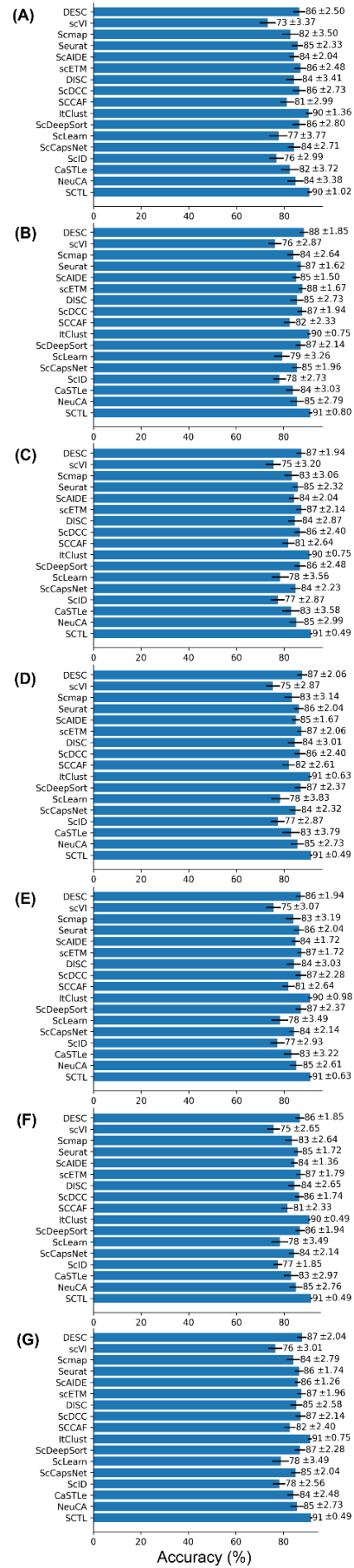

Fig. 3. Accuracy (%) of seventeen cell-type identification methods based on (A) Macosko et al. (B) Rheaume et al. (C) Tran et al. benchmark ocular datasets.

### 6.6.2 Accuracy of batch correction

We compared seventeen cell-type identification methods and evaluated the accuracy of the models when data has been generated based on several different batches using the Macosko et al. dataset. From seven different batches of this dataset, we assigned one single batch as testing dataset and the rest of the other batches as training dataset and reaped this process seven times to cover all seven batches as testing. Substantially smaller number of single cells exist in Batch B1 compared to the B2-B7 batches. Figure 4 (A-G) shows the accuracy (and standard error) of these methods based on batches B1 to B7, respectively.

Fig. 4. Accuracy (%) of seventeen cell-type identification methods based on seven different batches of Macosko et al dataset. We assigned each of batch B1 (A), B2 (B), B3 (C), B4 (D), B5 (E), B6 (F), B7 (G) as the testing dataset and the rest of the other batches as the training dataset.

The benchmarking results on Macosko et al. dataset showed that SCTL ItClust generate the highest accuracy in comparison to the other methods based on different batches. Both methods use a transfer learning-based algorithm, which seems better in removing the batch effect. The accuracy of models based on the B1 batch is lower compared to the other batches (see Figure 4A) which may suggest lower quality of cells in this batch.

### 6.6.3 Running time and peak memory usage

We assessed the computational time (Figure 5A) and peak memory usage (Figure 5B) for cell-type identification methods based on different number of cells (sample sizes) randomly selected from the Macosko et al. dataset.

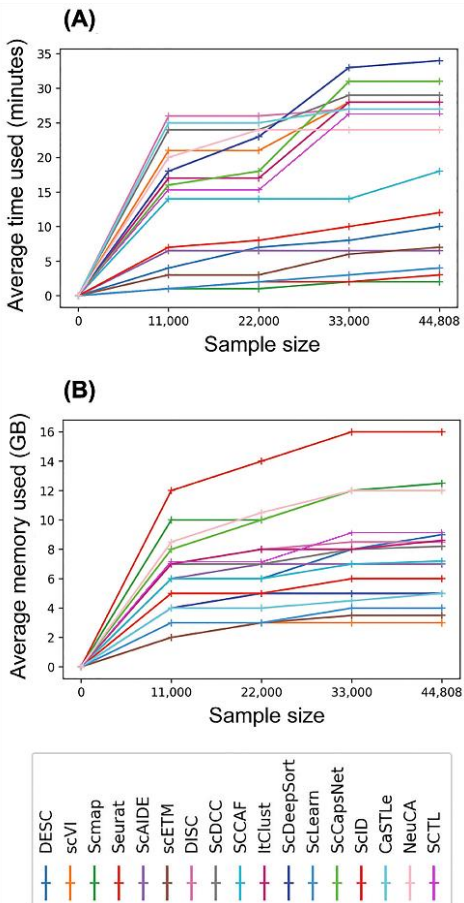

Fig. 5. Benchmarking cell-types identification methods. (A) Average memory used (GB) of sixteen cell-type identification methods on the Macosko et al. dataset. (B) Average time used (minutes) of sixteen cell-type identification methods based on the Macosko et al. dataset.

The experiments were performed on a workstation with 16 CPU cores at 3.90GHz, 64GB of memory and one GEFORCE RTX 3070 graphical processing unit (GPU) card.

As expected, almost all of the neural network-based methods (for example, ScDeepSort, ScCapsNet) need more computational time than the other methods, since a large number of parameters are required to be optimized. Generally, unsupervised cell-type identification methods show excellent performance in terms of speed and almost all of these methods require no more than 5 minutes to complete cell type or subtype identification.

To compare the computational time of the cell-type identification methods and to see how they are scaled when the number of cells increases, we randomly selected subsets of cell from the Macosko et al. dataset and benchmarked the methods. Results showed that the increase in computational time of Scmap and ScAIDE methods is smooth, suggesting that computational time increases almost linearly with increasing the sample size. Figure 5B shows the performance of cell-types identification methods based on computational time.

To compare the peak memory usage of the cell-type identification methods, we randomly selected subsets of cells from the Macosko et al. dataset and compared the methods. Results indicated that the scVI, scETM, and DESC required least amount of memory to complete the analyses.

Collectively, results suggest that applying cell-type identification methods on large datasets require larger memory sizes and longer run time thus the optimum size and computational time may be optimized based on the estimated numbers in Figure 5. Generally speaking, while supervised cell-type identification methods may lead to higher accuracy in comparison to unsupervised and semi-supervised methods, these methods require more computational resources.

## 7 CONCLUSION

In this study, we reviewed different single cell studies conducted in in vision science. We described numerous vision-related scRNA-seq/snRNA-seq atlases and provided taxonomies based on different species, tissue, and ocular diseases. We also described and benchmarked different machine-learning-based models in single cell data analysis and categorized models into three main subgroups: unsupervised_, semi-supervised_, and supervised_ cell-type and subtype identification methods. We also provided insights to scRNA-seq data visualization and provided several approaches for effectively visualizing the outcome in single cell studies. Finally, we benchmarked seventeen single cell type and subtype identification methods based on three different datasets encompassing diverse single retinal cell types and sub-types. We provided metrics on the accuracy, computational time, and memory usage thus allowing vision researchers to optimize their available computational resources when working with single cell transcriptomic datasets Our study provides a valuable review on available vision-related single cell datasets and scRNA-seq data analysis techniques and discusses future development of scRNA-seq based cell-type identification methods in vision science.

## ACKNOWLEDGMENT

This work was supported in part by grants from the Bright Focus Foundation.

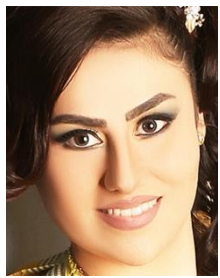

**Yeganeh Madadi** received her Ph.D. in Computer Engineering, Artificial Intelligence from the Azad University of Tehran in 2020 and her MSc in Computer Science from the Amirkabir University of Technology in 2015. Dr. Madadi is a Postdoctoral Research Fellow at the University of Tennessee Health Science Center. She was an Artificial Intelligence researcher at Aalborg University from 2019-2020. In addition, she has over 17 years of the executive experiment at the University of Tehran in Computer Engineering. Her research interests are Machine Learning, Computer Vision, and Bioinformatics.

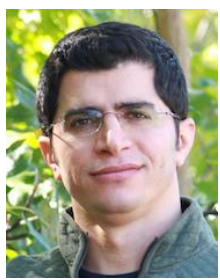

**Aboozar Monavarfeshani** earned his Ph.D. in Biological Sciences from Virginia Tech, and he is currently completing his Postdoctoral training at Boston Children's Hospital and the Center for Brain Science at Harvard University. During his doctoral training, he investigated molecules that are important for the formation of neural connections between retinal ganglion cells (RGCs) and neurons in several subcortical visual centers in the brain including the dorsal and ventral lateral geniculate nucleus, superior colliculus, and suprachiasmatic nucleus. In his Postdoc, Dr. Monavarfeshani is using single nucleus RNA sequencing to measure, both in post-mortem human tissues and in model organisms, the transcriptional changes that occur in degenerating RGCs, and in other cell types residing in the optic nerve head (ONH)–the primary site of glaucomatous injury to the axons of RGCs.

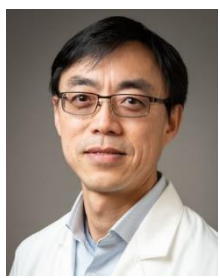

**Hao Chen** received his Ph.D. degree in Anatomy from Michigan State University. He is a full professor in the department of pharmacology, addiction science and toxicology at the University of Tennessee Health Science Center. Professor Chen has a wide range of research interests, with a focus on genetic and genomics of substance abuse related phenotypes modeled using rats. In addition, his lab contributed to many transcriptome studies of brain regions critically involved in the reward circuitry. Lastly, his lab is working on using deep learning to analyze rat social behaviors.

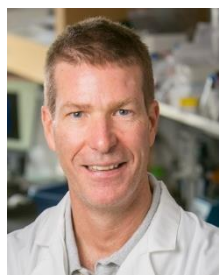

**W. Daniel Stamer,** Ph.D. was educated at the University of Arizona, earning his Bachelor of Science in Molecular and Cellular Biology in 1990 and doctorate in Pharmacology and Toxicology in 1996. After completing two research fellowships, Professor Stamer started his research program in 1998 at the University of Arizona, where he remained for 13 years; rising through the ranks to full Professor and Director of Ophthalmic Research. He was subsequently recruited to Duke University in 2011, where he currently serves as the Joseph A.C. Wadsworth Professor of Ophthalmology and Professor of Biomedical Engineering. The primary research focus of the Stamer laboratory is to understand the molecular and cellular mechanisms that regulate conventional outflow such that novel targets can be identified, validated and used for the development of therapeutics that target/modify the diseased tissue responsible for elevated intraocular pressure in glaucoma. Over the past 30+ years, Professor Stamer has pioneered the development of cellular, tissue, organ culture and murine model systems for use by his laboratory and others to study conventional outflow physiology and pharmacology. His laboratory has worked closely with industry, assisting in the development/pre-clinical testing of several new classes of glaucoma drugs that target the diseased conventional outflow pathway responsible for ocular hypertension. Research progress is documented in over 175 peer-reviewed primary contributions to the literature and two dozen reviews/book chapters/editorials, having over 9900 citations. His work was recognized by the Rudin Prize for Glaucoma in 2012 and the Research to Prevent Blindness Foundation in 2013. More recently, Professor Stamer was elected as ARVO trustee in 2015 and ARVO president in 2019/20 and elected into the Glaucoma Research Society in 2022. He currently holds prominent editorial positions in three premier ophthalmology journals: the Journal of Ocular Pharmacology and Therapeutics, Investigative Ophthalmology and Visual Science and Experimental Eye Research. Moreover, Professor Stamer currently serves on the scientific advisory boards for 6 companies and three foundations that support glaucoma research.

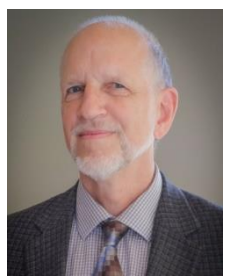

**Robert (Rob) W. Williams** received a Bachelor of Science in neuroscience from UC Santa Cruz (1975) and a Ph.D. in physiology at UC Davis with Leo M. Chalupa (1983). He did postdoctoral work in developmental neurobiology at Yale with Pasko Rakic and moved to the University of Tennessee in 1989. Professor Williams is chair of the Department of Genetics, Genomics and Informatics and holds the UT Oak Ridge National Laboratory Governor's Chair in Computational Genomics. He was president of the International Society for Behavioral and Neural Genetics and is founding director of the Complex Trait Community (www.complextrait.org). He was editor-in-chief of Frontiers in Neurogenomics for over a decade, and serves on the editorial boards of Genes, Brain & Behavior, EBM, Neuroinformatics, Mammalian Genome, Molecular Vision, Alcohol, BiomedCentral Neuroscience, the Journal of Biomedical Discovery and Collaboration, and Behavior Genetics. Research progress is documented in over 25690 citations. One of Prof. Williams' more notable contributions is in the field of systems genetics and expression genetics. He and his research group have built GeneNetwork (www.genenetwork.org), an online resource and suite of phenotype and genotype data and analysis code that is used widely by the genetics and molecular biology communities.

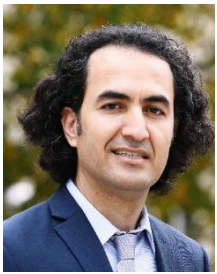

**Siamak Yousefi** is an Assistant Professor at the Department of Ophthalmology and the Department of Genetics, Genomics, and Informatics of the University of Tennessee Health Science Center (UTHSC) in Memphis. He received his PhD in Electrical Engineering from the University of Texas at Dallas in 2012 and completed two postdoctoral trainings at the University of California Los Angeles (UCLA) and University of California San Diego (UCSD). He is the director of the Data Mining and Machine Learning (DM2L) laboratory at UTHSC working on broad applications of Artificial Intelligence (AI) in vision and ophthalmology particularly glaucoma diagnosis, prognosis, and monitoring. Dr. Yousefi is a senior member of the IEEE and a member of ARVO.